\documentclass[letterpaper, 10 pt, conference]{ieeeconf}  

\IEEEoverridecommandlockouts                              
\usepackage{cite}
\usepackage{amsmath,amssymb,amsfonts}
\usepackage{enumerate}

\usepackage{algorithmic}
\usepackage{graphicx}
\usepackage{subfigure}
\usepackage{textcomp}
\usepackage{comment}
\usepackage{xcolor}
\usepackage{makecell} 
\usepackage{multirow} 

\title{\LARGE \bf
Deep Transfer Learning for System Identification Using Long Short-Term Memory Neural Networks*
}

\author{Kaicheng Niu$^1$, Mi Zhou$^2$, Chaouki Abdallah$^3$, and Mohammad Hayajneh$^4$
\thanks{*This work was not supported by any organization}
\thanks{$^1$C. Abdallah is with School of Electrical and Computer Engineering,
        Georgia Institute of Technology, North Avenue Northwest, Atlanta, GA 30332, USA 
        {\tt\small ctabdallah@gatech.edu}}%
\thanks{$^2$K. Niu is with School of Electrical and Computer Engineering, 
		Georgia Institute of Technology, North Avenue Northwest, Atlanta, GA 30332, USA 
        {\tt\small kniu9@gatech.edu}}%
\thanks{$^3$M. Zhou is with School of Electrical and Computer Engineering, 
		Georgia Institute of Technology, North Avenue Northwest, Atlanta, GA 30332, USA 
        {\tt\small mizhou2018@gatech.edu}}%
\thanks{$^4$M. Hayajneh is with College of Information Technology, United Arab Emirates University, Sheik Khalifa Bin Zayed Street, P. O. Box 17551, Al-Ain, UAE
		{\tt\small mhayajneh@uaeu.ac.ae}}
}

\begin{document}

\maketitle
\thispagestyle{empty}
\pagestyle{empty}

\begin{abstract}
Recurrent neural networks (RNNs) have many advantages over more traditional system identification techniques. They may be applied to linear and nonlinear systems, and they require fewer modeling assumptions. However, these neural network models may also need larger amounts of data to learn and generalize. Furthermore, neural networks training is a time-consuming process. Hence, building upon long-short term memory neural networks (LSTM), this paper proposes using two types of deep transfer learning, namely parameter fine-tuning and freezing, to reduce the data and computation requirements for system identification. We apply these techniques to identify two dynamical systems, namely a second-order linear system and a Wiener-Hammerstein nonlinear system. Results show that compared with direct learning, our method accelerates learning by 10\% to 50\%, which also saves data and computing resources.

\end{abstract}

\section{INTRODUCTION}\label{sec:intro}

Control and system theory have matured in the past few decades. In addition, with the availability of high-performance computers and advanced algorithms, engineers are now able to manipulate many complex dynamical systems. Before control algorithms are applied however, an important step is to understand the system and its corresponding dynamics. For example, if one hopes to control an unmanned aerial vehicle (UAV), it is critical to model how the UAV responses to thrusters. As a consequence, the problem of \textit{system identification} became an important research area in system theory.%

The system identification problem has many formulations in different contexts, with a widely-accepted one proposed by L. Ljung (1978) \cite{b1}. Ljung described identification as a process to select a mathematical model from a set of model candidates in order to best fit the system input-output behavior with respect to a given criterion. We will use Ljung's formulation in the remainder of our paper.%

Ljung's definition of system identification emphasized three key elements, namely the data (both inputs and outputs), the model set and the evaluation criterion. The first two elements are often ad-hoc and vary from one application to another. On the other hand, there exist some commonly used performance criteria, including the \textit{least-squares criterion} \cite{b2} used here.  We thus propose a mathematical model that minimizes the mean squared error (MSE). 

More specifically, suppose we have a discrete system $\mathcal{S}$ and an identification model $\mathcal{\hat{S}}$. The inputs to both systems are $\mathbf{U} = \{\mathbf{u}_i\}_{i=0}^{T-1}$. If the outputs of $\mathcal{S}$ are $\mathbf{Y} \triangleq \{\mathbf{y}_i\}_{i=0}^{T-1}$ and the outputs of $\mathcal{\hat{S}}$ are $\mathbf{\hat{Y}} \triangleq \{\mathbf{\hat{y}}_i\}_{i=0}^{T-1}$, then the MSE for this prediction is:
\begin{equation}
	\label{e1}
	l(\mathbf{Y}, \mathbf{\hat{Y}}; \mathcal{S}, \mathcal{\hat{S}}, \mathbf{U}) \triangleq \dfrac{1}{T}\sum_{i=0}^{T-1}\Vert \mathbf{y}_i - \mathbf{\hat{y}}_i\Vert_2^2. 
\end{equation}

Note that \eqref{e1} may be easily extended for multiple inputs and outputs. Eqn. \eqref{e1} will be used as the identification criterion for this paper.  Our identification problem is to find a model $\mathcal{\hat{S}}$ in our model set to minimize error \eqref{e1}.

Eqn. \eqref{e1} reframes the system identification problem into an optimization problem, whose solutions are abundant in the literature. Traditional methods include \cite{b1, b2,b11} and so on. Recently, with the rapid growth of computers, identifying a system using machine learning and deep learning algorithms has become feasible and practical. Early work by Chen, Billings and Grant (1990) \cite{b3}, attempted to predict the output of a nonlinear system based on a multiple-layer neural network. In 1998, Lu and Basar \cite{b4} introduced radial basis function network to identify nonlinear systems. Delgado, Kambhampati and Warwick (1995) \cite{b5} applied recurrent neural networks (RNN) for nonlinear-control affine systems. More recently, the long short-term memory (LSTM) neural network, first proposed by S. Hochreiter and J. Schmidhuber (1997) \cite{b19} and developed by Gers, Schmidhuber and Cummins (2000) \cite{b6}, has drawn great attention. Building upon LSTM, many practical neural network structures have been proposed for identification \cite{b7} \cite{b8}.%

Neural networks, especially recurrent neural networks, have many advantages in system identification problems. For example, a neural network may be applied to linear and non-linear systems. We can also assume little \textit{a priori knowledge} of the system structure before using this approach. The disadvantages of this approach are also obvious. Firstly, training an RNN requires a large amount of data sampled from the target system. Secondly, training neural network is time-consuming. Finally, a neural network only works on the designated system and if the system were to change, either in parameter or in structure, the network model may no longer work.%

To address these problems, this paper applies \textit{deep transfer learning} for system identification. As indicated by its name, deep transfer learning is the realization of transfer learning using \textit{deep neural networks}. Our work exploits two different techniques of deep transfer learning, namely \textit{fine-tuned} and \textit{frozen} networks. In the first case, all parameters in the neural network are updated, while in the second case, parameters in some of the network layers are frozen. For more information on deep transfer learning, please refer to the survey papers \cite{b9}, \cite{b12} and \cite{b13}. The main contribution of our work is to accelerate the training of neural networks for system identification and to make it easier to adapt an existing model to a new identification problem.%

The remainder of this paper is organized as follows: in Section \ref{sec:dynamics}, we review basic ideas in systems identification. Section \ref{sec:NN_TL} describes neural networks and transfer learning, specifically RNN, LSTM and deep transfer learning. Section \ref{sec:DTL} explains the use of deep transfer learning in system identification problems, including the design of neural networks and the process of transfer learning. Based on these methods, Section \ref{sec:simulation} presents a series of experiments on different dynamical systems. Section \ref{sec:conclusion} summarizes our conclusions.%

\section{Dynamical Systems and Identification} \label{sec:dynamics}
In this section, we review concepts in dynamical systems and system identification. Generally speaking, a dynamical system may be expressed by a \textit{state space model} \cite{b14}:
\begin{equation}
	\label{e2}
	\begin{split}
		\mathbf{\dot{x}} &= \mathbf{f}(t, \mathbf{x}, \mathbf{u})\\
		\mathbf{y} &= \mathbf{h}(t, \mathbf{x}, \mathbf{u}),
	\end{split}
\end{equation}
where $\mathbf{x} \in \mathbb{R}^N$ is the \textit{state} of the system; $\mathbf{u}\in\mathbb{R}^M$ is the input, and $t$ is the time. The map $\mathbf{f}: \mathbb{R}^{1 + N + M} \rightarrow \mathbb{R}^N$ describes how the state evolves, and $\mathbf{h}: \mathbb{R}^{1 + N + M}\rightarrow \mathbb{R}^P$ maps time, state and input to output.%

If the system is a discrete-time system, it is represented by the following equations:
\begin{equation}
	\label{e3}
	\begin{split}
		\mathbf{x}_{n+1} &= \mathbf{f}(n,\mathbf{x}_n, \mathbf{u}_{n})\\
		\mathbf{y}_{n} &= \mathbf{h}(n, \mathbf{x}_n, \mathbf{u}_n),
	\end{split}
\end{equation}
where $n\in \mathbb{Z}$ is the time index. Our work focuses on discrete-time systems. One special class of discrete-time systems are discrete-time, time-invariant linear systems \cite{b15}:
\begin{equation}
	\label{e4}
	\begin{split}
		\mathbf{x}_{n+1} &= \mathbf{Ax}_n + \mathbf{Bu}_n\\
		\mathbf{y}_{n} &= \mathbf{Cx}_n + \mathbf{Du}_n,
	\end{split}
\end{equation}
where $\mathbf{A}\in\mathbb{R}^{N\times N}$, $\mathbf{B}\in\mathbb{R}^{N\times M}$, $\mathbf{C}\in\mathbb{R}^{P\times N}$ and $\mathbf{D}\in\mathbb{R}^{P\times M}$. $\mathbf{y}_n$ is the $n$-th output, $\mathbf{x}_n$ is the $n$-th state and $\mathbf{u}_n$ is the $n$-th input.%

Another special case is that of nonlinear time-invariant systems:

\begin{equation}
	\label{e5}
	\begin{split}
		\mathbf{x}_{n+1} &= \mathbf{f}(\mathbf{x}_n, \mathbf{u}_{n})\\
		\mathbf{y}_{n} &= \mathbf{h}(\mathbf{x}_n, \mathbf{u}_n).
	\end{split}
\end{equation}

\eqref{e4} and \eqref{e5} are the systems we will be dealing with.

As stated in Section \ref{sec:intro}, our work mainly focuses on minimizing the MSE between the actual system outputs and the identified model outputs. The expression of MSE is shown in Eqn. \eqref{e1}. In our setting, the identified system is modeled as the LSTM neural network.

\section{Neural network and transfer learning} \label{sec:NN_TL}
\subsection{Neural Network}
\begin{figure}[htbp]
	\begin{center}
			\includegraphics[width = 15pc]{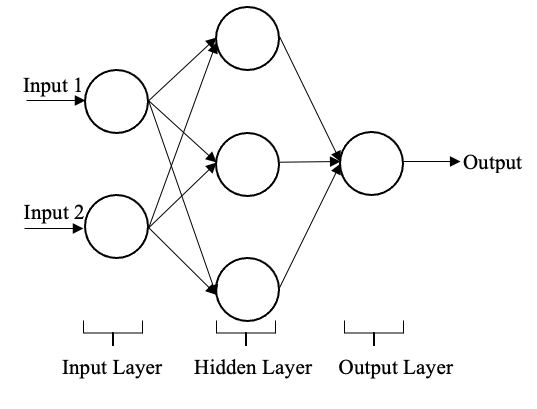}
	\end{center}
	\caption{An example of feed-forward network}
	\label{fig2}
\end{figure}

We start from the \textit{neuron}, the smallest unit in a neural network. According to Nielsen (2015) \cite{b16}, a neuron is a mathematical model that takes $N$ inputs and yields $1$ output. The relationship between the inputs and the output can be described as Eqn. \eqref{e6},
\begin{equation}
	\label{e6}
	y = \sigma(\mathbf{w}^\mathrm{T}\mathbf{x} + b),
\end{equation}
where $\mathbf{x}\in \mathbb{R}^N$ is a column vector consisting of $N$ inputs; $\mathbf{w} \in \mathbb{R}^{N}$ describes the \textit{weights} of these inputs. $b \in \mathbb{R}$ is a constant \textit{bias} vector. Function $\sigma: \mathbb{R} \rightarrow \mathbb{R}$ is the \textit{activation function} that maps the combination of weighted inputs into a single output. Arranging a number of neurons in parallel yields a single layer of a neural network. More complex neural networks may be obtained by simply serially connecting multiple layers. Fig. \ref{fig2} provides an example of a neural network with $3$ layers made up of $6$ neurons. This structure of neural networks is often referred to as \textit{feed-forward network} \cite{b16} or \textit{multiple layer perceptron} \cite{b17}. If there are more than one hidden layer in a neural network, the neural network is a \textit{deep neural network} (DNN) \cite{b16}.

Unlike traditional feed-forward neural networks, a \textit{recurrent neural network} (RNN) is characterized by the presence of feedback channels. If a neuron receives feedback signals from its output, the output of this neuron is written as:
\begin{equation}
	\label{e7}
	\begin{split}
		y_n = \sigma(wx_n + vy_{n-1}),
	\end{split}
\end{equation}
where $w$ and $v$ are weights corresponding to the current input $x_n$ and last output $y_{n-1}$. Note that a neuron can also receive feedback signals from other neurons by simply replacing $y_{n-1}$ with other outputs. For more information, please refer to \cite{b20}.

A special case of RNN is the \textit{long-short term memory neural network} (LSTM), first proposed in \cite{b19}. According to \cite{b19}, error signals in a standard RNN may explode or vanish after a long time period. In order to deal with this problem, LSTM introduced two extra structures, namely an \textit{input gate} and an \textit{output gate}, which retain long-time error signals. Our work on transfer learning will be based on the LSTM neural network.

\subsection{(Deep) Transfer Learning}
Transfer learning aims at improving the accuracy of a machine learning model for one problem (called the \textit{target}) by reusing its knowledge from another problem (called the \textit{source}). The data sets in the target or source problems are called \textit{target domain} or \textit{source domain}, while the learning tasks are called a \textit{target task} or \textit{source task}. Furthermore, if the machine learning model aforementioned is a deep neural network, this process is termed as deep transfer learning. Reference \cite{b9} provides exact definitions of transfer learning and deep learning together with their classifications. As mentioned earlier, we will explore two techniques for transfer learning: a fine-tuned neural network and a frozen neural network.
\subsubsection{Fine-tuned neural network} A transferred neural network is called fine-tuned if it is trained directly on the target domain after being trained on the source domain. Before fine-tuning, all the parameters (or weights of the neural network) from the source domain are kept. When transferred into the target domain, these parameters will be further adapted to the new problem. References \cite{b23} and \cite{b24} provide successful applications of fine-tuned neural networks.
\subsubsection{Frozen neural network} Unlike the fine-tuned case, when transfer learning using a frozen neural network, we fix some parameters learned from the source domain before reusing them. We may also completely abandon some parameters from the source and replace them with new ones.

\section{Deep Transfer Learning for Identification} \label{sec:DTL}
We now frame the problem of deep transfer learning for identification problems. This section discusses four components, including the dataset, the neural network structure, the application of transfer and the training and testing.
\subsection{Dataset}
As mentioned above, in order to identify a system, we first gather input-output data. For transfer learning, we need input and output data from both the source and target systems. 
\subsubsection{System input (feature)} When training a neural network, the system inputs,  also called \textit{features} are presented to the first layer. Oftentimes, inputs are random sequences selected from a specified distribution. Commonly used distributions include the Gaussian distribution, the uniform distribution, the truncated normal distribution and so on. Since neural networks prefer normalized data, or in other words, data that are concentrated into a small range like $[-1,1]$, we select our input data from a truncated normal distribution. According to \cite{b25}, the probability density function (PDF) of a truncated normal distribution defined on $[a,b]$ with mean $\mu$ and standard deviation $\sigma$ is given by Eqn. \eqref{e8}.
\begin{equation}
	\label{e8}
	\Psi(x; \mu, \sigma^2, a, b) =  \left\{
	\begin{aligned}
		&0,\ x \leq a\\
		&\dfrac{\phi(x; \mu, \sigma^2)}{\Phi(b; \mu, \sigma^2) - \Phi(a; \mu, \sigma^2)},\ a < x < b \\
		&0, \  b \leq x
	\end{aligned}
	\right.,
\end{equation}
where $\phi$ represents the PDF of normal distribution; $\Phi$ is the cumulative distribution function (CDF) of a normal distribution. We will use $a = -1$, $b = 1$, $\mu = 0$, $\sigma = 1$ in our experiments. 

\subsubsection{System output (label)} The system output is obtained by driving the system with the inputs described above. Since this output is used to train the LSTM network, it is also called the \textit{label} of the data. During our simulation, we keep the initial state of the system at $0$. 

Our experimental dataset is divided into two parts: a \textit{training set} and a \textit{test set}. The training set is used to train the model and obtain the neural networks weights. Each output sequence in the training set has a length of $5,000$. The test set is used to evaluate the performance of the model and has a length of $10,000$. Note that the length of the test set is larger than that of the training set, which is used to verify the generalization and prediction abilities of the trained model. 
For the purpose of evaluation and benchmarking, we sample $160$ input and output sequences, divided into $5$ groups, with $32$ sequences in each group. These 32 sequences form a mini-batch. However, not all data are used up in our experiments, and in practical projects and experiments, one may not be able to gather such an amount of data.

\subsection{Neural Network Structure}
Now we consider the structure of our neural network model.
\subsubsection{Input layer} Our neural network takes in two inputs. The first input is the system input as described in the last subsection while the second input is the sample index normalized to the range $[0, 1]$. For example, if we have $100$ inputs in total, from $0$ to $99$, the normalized index will be $0$, $0.01$, ..., $0.99$. The reason the index is included is because during our experiments, we found that the learning became more robust when fed with the index. These two inputs are concatenated together with a concatenate layer, forming the final input to the model.
\subsubsection{LSTM layer} The LSTM layers make up the most important part of our model. We use three LSTM layers that are connected sequentially. They have 16, 64 and 128 LSTM neurons, respectively.

\subsubsection{Output layer} The output LSTM layer has a length of 128, yet the output of the dynamic system has length 1. Hence, an ordinary neural network layer, called a dense layer, is required for dimension matching. 

\begin{figure}[htbp]
	\includegraphics[width = 21pc]{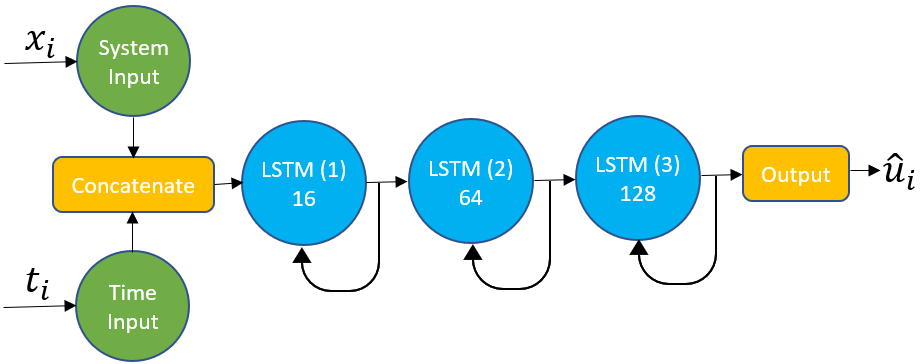}
	\caption{LSTM neural network structure}
	\label{fig7}
\end{figure}

Fig. \ref{fig7} shows the structure of our neural network.

\subsection{Applying Transfer Learning}
As mentioned above, we will use both fine-tuning and freezing for transfer learning. For fine-tuning, we first train the neural network on the source domain, then continue to train it on the target domain. After a number of experiments and parameter adjusting, we decided to train it on the source domain for $10$ epochs and then train it on the target domain until convergence is achieved. For the freezing case, we train the neural network on the source domain, keep the weights in the first two LSTM layers (LSTM1 and LSTM2) frozen, and replace the others with new ones initialized with random values, then tarin them on the target domain. For the frozen layers, we train the model on the source domain for $40$ epochs.

\subsection{Training and Testing}
We now examine the training and testing stages of the neural network, which is the last preparation step in our experiment. 
\subsubsection{Training} The training of the LSTM is based on the Adam optimizer \cite{b26}. Here we use default step size $0.01$. The stopping criterion is selected as:
\begin{equation}
	\vert e_{n} - e_{n-1} \vert < 5 \times 10^{-5},
\end{equation}
where $e_{n}$ is the MSE after the $n$-th epoch of training. If the difference between two iterations is small, we assume that the model has reached a local minimum and thus stop our learning.

The training data are used as follows: with each group (or each \textit{batch}) of data mentioned above, we train the neural network for $10$ epochs, then move to another batch. By doing so, the generalization ability of the neural network is guaranteed and overfitting is reduced. Note again, we have significantly more data than required for the purpose of experiment.

\subsubsection{Testing} We test the model with a data length of $10,000$, in order to evaluate the generalization and long-term prediction abilities of the model. The testing set is a mini-batch with $32$ sequences of data, so the dimension of test set is $32\times 10,000$. The mean squared error is calculated by averaging the MSE on each data point.

Unlike traditional applications, where only one neural network is used and sufficient data are presented, the aim of our approach is to study how the transferred neural network performs with insufficient data. Hence, under this scenario, we need to set up metrics to compare the performance of two different neural networks. Here, we basically monitor two criteria:
\begin{enumerate}
	\item \textit{Constant MSE}: how many training epochs are required to reach an MSE of $1 \times 10^{-2}$;
	\item \textit{Dynamic MSE}: how many training epochs are required to reach \textit{two times of minimal MSE}.
\end{enumerate}

Criterion 1) is easier to understand, because in many applications that do not have a strict requirement of accuracy, $1 \times 10^{-2}$ MSE is a good estimate of a "small" error.

\begin{figure}[htbp]

		\includegraphics[width = 21pc, height = 8pc]{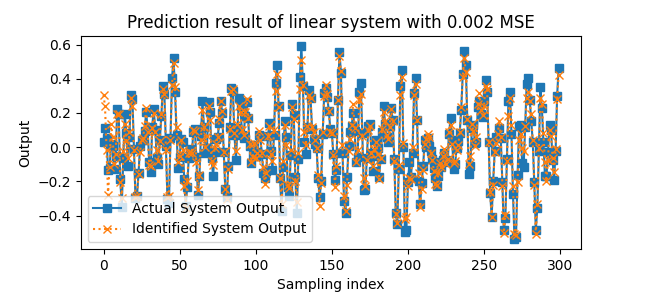}
	
	\caption{
		Prediction result of linear system when MSE is 0.002
	}
	\label{fig9}
	\vspace{-10pt}
\end{figure}

The reason why we set criterion 2) is because, with limited data and/or insufficient training epochs, we would expect the neural network will perform worse than the regular case. From this perspective, twice the minimal MSE is a reasonable compromise. During our experiment, we note that the MSE is around $1\times 10^{-3}$ when the algorithm converges, hence the dynamic MSE is around $2 \times 10^{-3}$, which gives a better result than constant MSE. Fig. \ref{fig9} illustrates identification result of a linear system when the MSE is 0.002.

\section{Numerical experiments}\label{sec:simulation}
In this section, we perform transfer learning techniques for both linear and nonlinear systems. For the linear case, we choose a second-order and a third-order system, and transfer from the third-order system to the second-order system. For the nonlinear case, we use a Wiener-Hammerstein benchmark system as proposed in \cite{b21}. All the experiments are conducted using Python 3.6.1 and TensorFlow 2.3. To make our experiments repeatable, we set the random seed to $2021$.
\subsection{Linear System Identification}
Consider a single-input single-output (SISO) third-order linear, time-invariant system given by \eqref{eq5_1}:
\begin{equation}
	\label{eq5_1}
	\begin{split}
		\mathbf{x}_{n+1} &= \begin{bmatrix}
		0.60&0.00&0.00\\
		0.70&0.15&-0.80\\
		0.45&0.80&0.45
		\end{bmatrix}\mathbf{x}_n + \begin{bmatrix}
		1.60\\0.70\\0.50
 	\end{bmatrix}	\mathbf{u}_n\\
 	\mathbf{y}_n &= \begin{bmatrix}
 	0.05 & 0.10 & 0.20
 \end{bmatrix}\mathbf{x}_n + \begin{bmatrix}
 	0.01
 \end{bmatrix}\mathbf{u}_n .
	\end{split}
\end{equation}
We will use this system as the original problem (the source) for our transfer learning. As stated before, the system is driven by a truncated normal distribution with the support $[-1, 1]$. We also create a second-order linear system as the target for identification, given by Eqn. \eqref{eq5_2}. We assume the initial state of the system is zero for convenience. 

\begin{equation}
	\label{eq5_2}
	\begin{split}
		\mathbf{x}_{n+1} &= \begin{bmatrix}
		0.20&-0.70\\
		0.70&0.50
		\end{bmatrix}\mathbf{x}_n + \begin{bmatrix}
		1.00\\0.70
 	\end{bmatrix}	\mathbf{u}_n\\
 	\mathbf{y}_n &= \begin{bmatrix}
 	0.10 & 0.25
 \end{bmatrix}\mathbf{x}_n + \begin{bmatrix}
 	0.15
 \end{bmatrix}\mathbf{u}_n. 
	\end{split}
\end{equation}

\begin{figure}[htbp]
	\includegraphics[width = 21pc, height = 15pc]{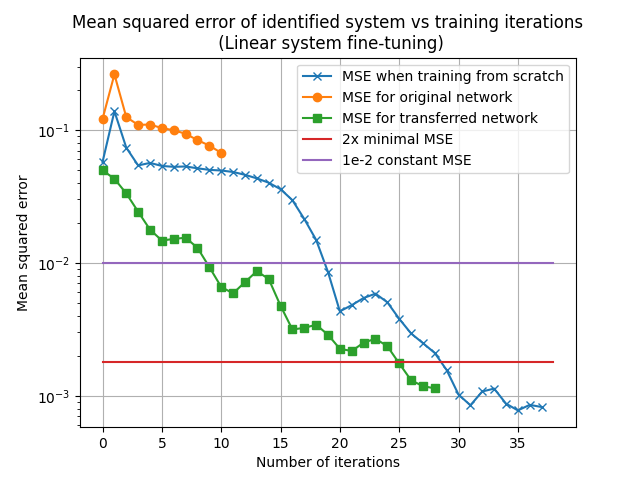}
	\caption{Identified linear system MSE vs the number of training iterations for raw network and fine-tuned neural network}
	\label{fig5_3}
\end{figure}

We first implement transfer learning using fine-tuning. We train the neural network on system \eqref{eq5_1} for 10 epochs, then train it on the target system \eqref{eq5_2} until convergence is achieved. The training result is shown in Fig. \ref{fig5_3}. The blue line indicates the mean-squared error for the "standard" neural network, or in other words, the mean squared error when training the network from scratch. The orange line indicates the mean-squared error trained on the source domain, system \eqref{eq5_1}. The green line is the MSE for fine-tuned network on the target domain, namely system \eqref{eq5_2}.%

The result from this experiment shows that the neural network trained from scratch converged after $37$ epochs, and the transferred network converged after $28$ epochs. It took the raw neural network $19$ iterations to reach an MSE of $1\times 10^{-2}$, while for transferred neural network, convergence was reached after $9$ epochs, which was $52.6\%$ lower. Also, in this case, the dynamic MSE ($1.783\times 10^{-3}$) was reached by the raw neural network after $29$ epochs and by the transferred neural network after $25$ epochs, which was $13.8\%$ lower.%

\vspace{-10pt}

\begin{figure}[htbp]
	\includegraphics[width = 21pc, height = 15pc]{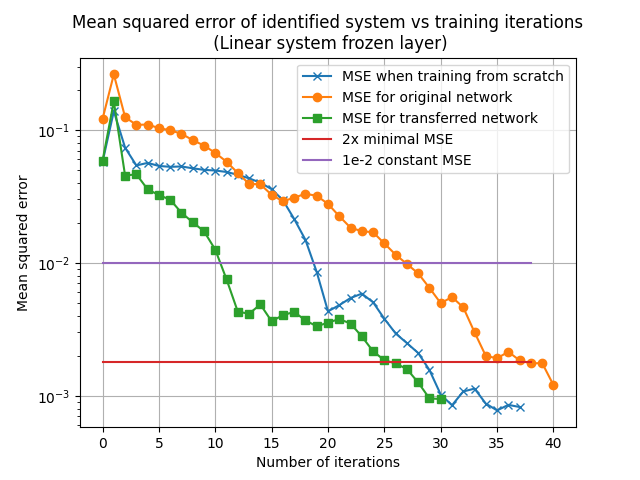}
	\caption{Identified linear system MSE vs the number of training iterations for raw network and neural network with frozen layers}
	\label{fig5_4}
\end{figure}

Next we performed transfer learning with frozen layers. The frozen neural network requires more epochs to obtain a prior knowledge on the source domain, and we trained it to identify system \eqref{eq5_1} for 40 epochs. The result in Fig. \ref{fig5_4} shows that it takes the transferred neural network $42.1\%$ less epochs to reach an MSE of $1\times 10^{-2}$ and $10.3\%$ less epochs to reach dynamic MSE, namely $1.783\times 10^{-3}$ (same as before).

\vspace{-10pt}

\subsection{Wiener-Hammerstein System}
\begin{figure}[htbp]
	\includegraphics[width = 21pc]{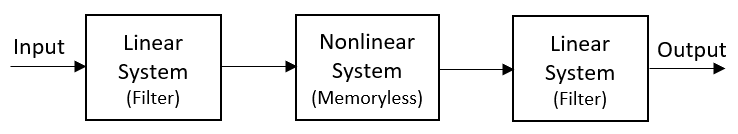}
	\caption{A typical structure of Wiener-Hammerstein system}
	\label{fig5_5}
\end{figure}

The Wiener-Hammerstein system is a combination of a Wiener and a Hammerstein systems. One general case of Wiener-Hammerstein system consists of three interconnected subsystems, starting with a linear filter, followed by a nonlinear system without memory, and ending with another linear filter\cite{b22}. This structure is shown in Fig. \ref{fig5_5}.

 Our experiment follows the benchmark proposed in \cite{b21}, which is organized as follows: the first subsytem is a third-order Chebyshev filter, the nonlinearity in the second part is introduced by a diode and the last subsytem is another Chebyshev filter. Note that \cite{b21} used a continuous-time model, but our system evolves in discrete time. The first Chebyshev filter is given by:
\begin{equation}
 	\label{eq5_3}
 	\begin{split}
 		y_N = &a_0u_N + a_1u_{N-1} +  a_2u_{N-2} + a_3u_{N-3}\\ &+b_1y_{N-1} + b_2y_{N-2} + b_3y_{N-3},
 	\end{split}
\end{equation}
where $a_0 = 0.0083$, $a_1= 0.0248$, $a_2 = 0.0248$, $a_3 = 0.0083$,
$b_1 = 2.2800$, $b_2 = 1.9766$ and $b_3 = 0.6307$. The second Chebyshev filter takes the same form as Eqn. \eqref{eq5_3}, with a different set of parameters: $a_0 = 0.7452$, $a_1= 1.3902$, $a_2 = 1.3902$, $a_3 = 0.7452$,
$b_1 = -1.4250$, $b_2 = -1.2920$ and $b_3 = -0.5538$. For the nonlinear system, we use the slightly-modified model to that in \cite{b21},
\begin{equation}
		f(x) = \left\{
	\begin{aligned}
		&\dfrac{10}{11}x,\ x < 0\\
		&x,\ 0 \leq x \leq \dfrac{3}{10} \\
		&\dfrac{3}{10}, \  \dfrac{3}{10} < x
	\end{aligned}
	\right..
\end{equation}

Now we introduce our original system in the source domain as a second-order Chebyshev filter, shown in Eqn.~\eqref{eq5_4}:
\begin{equation}
	\label{eq5_4}
	y_N = a_0u_N + a_1u_{N-1} +  a_2u_{N-2} + b_1y_{N-1} + b_2y_{N-2}.
\end{equation}
Here $a_0 = 0.0635$, $a_1 = 0.1270$, $a_2 = 0.0635$, $b_1 = 1.2129$ and $b_2 = -0.6646$.

\begin{figure}[htbp]
 	\includegraphics[width = 21pc, height = 15pc]{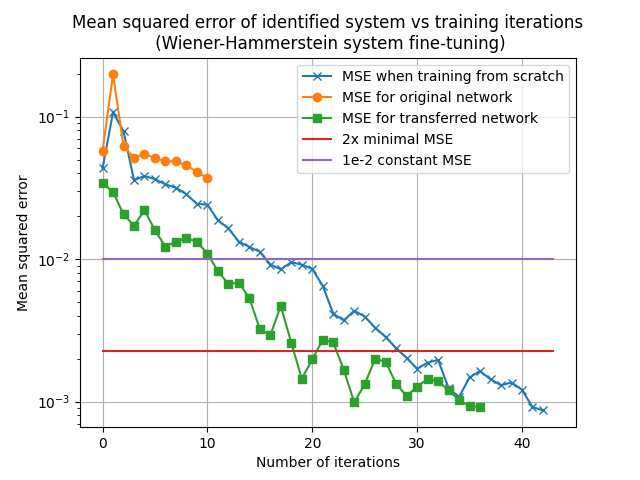}
 	\caption{Identified Wiener-Hammerstein system MSE vs the number of training iterations for raw network and fine-tuned neural network}
 	\label{fig5_9} 	
 	\vspace{-10pt}
 \end{figure}

The result of fine-tuning tarnsfer learning is shown in Fig.~\ref{fig5_9}. It took $31.3\%$ less iterations for the transferred neural network to reach $1\times 10^{-2}$ MSE and $34.5\%$ less to reach dynamic MSE, which is $2.264\times 10^{-3}$.
 
\begin{figure}[htbp]
	\includegraphics[width = 21pc, height= 15pc]{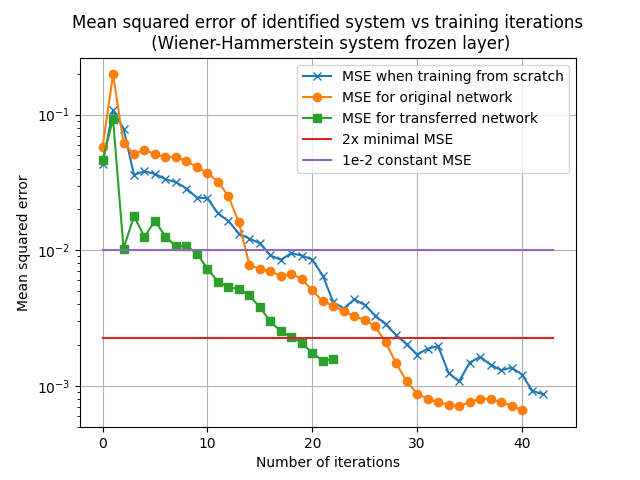}
	\caption{Identified Wiener-Hammerstein system MSE vs the number of training iterations for raw network and neural network with frozen layers}
	\label{fig5_10} 	
	\vspace{-20pt}
\end{figure}

The transferred neural network with frozen layers worked a little bit better than the fine-tuned network, as shown in Fig. \ref{fig5_10}. It reached $1\times 10^{-2}$ MSE with $43.75\%$ less training and reached dynamic MSE with $34.5\%$ less training. Note that here the dynamic MSE is still $2.264\times 10^{-3}$, which is the same as before.

During our experiment, we notice that transfer learning is especially effective if it is desired to reach a relatively high but still tolerable ($1\times 10^{-2}$ for instance) mean-squared error. Also, it is important to mention that during our experiment, we find that the MSEs of transferred network and raw network tend to be the same as the number of epochs gets larger (e.g., as large as 100). However, the more training you want, the more data will be needed, otherwise overfitting is likely to result. Hence, we note that transfer learning achieves a good balance between insufficient data and a demand of accuracy.

\section{conclusions and future works}\label{sec:conclusion}
In this paper, we propose deep transfer learning for system identification problems using LSTM, a special structure of recurrent neural networks. The main goal of this method is to accelerate the training of LSTM, which is especially meaningful when only a small amount of data is available. We perform experiments using both fine-tuning and freezing techniques. The results show that the number of epochs for training has significantly decreased under two different criteria.

Although this approach makes up for the lack of data, it has some shortcomings. For example, it requires many parameters that can only be determined empirically. Also, the selection of source domain (or the original system) has an impact on the transfer learning result. Hence, our further work will concentrate on finding out a generic method to determine the original system, such that the transfer learning result may be satisfactory for many commonly encountered dynamical systems. Finally, we will apply transfer learning identification methods for multiple input multiple output (MIMO) systems or time-variant (TV) systems.

\end{document}